\documentclass[twocolumn,showpacs,aps,prl]{revtex4}
\usepackage{epsfig,amssymb}

\begin{document}

\title{Normal-superfluid interaction dynamics in a spinor Bose gas}
\author{J.~M. McGuirk$^\ast$, D.~M. Harber, H.~J. Lewandowski$^\ast$, and E.~A. Cornell$^\ast$}
\affiliation{JILA, National Institute of Standards and Technology and Department of Physics, \\
University of Colorado, Boulder, Colorado 80309-0440}
\date{\today}

\begin{abstract}
Coherent behavior of spinor Bose-Einstein condensates is studied in the presence of a significant uncondensed (normal) component. Normal-superfluid exchange scattering leads to a near-perfect local alignment between the spin fields of the two components.  Through this spin locking, spin-domain formation in the condensate is vastly accelerated as the spin populations in the condensate are entrained by large-amplitude spin waves in the normal component. We present data evincing the normal-superfluid spin dynamics in this regime of complicated interdependent behavior.
\end{abstract}

\maketitle

A spinor Bose gas is a rich system for studying quantum coherence and the interaction dynamics of a superfluid with a normal component.  The complex spin dynamics of such a system exhibit behavior that is not seen in either component individually.  Mean field effects, exchange scattering, and elastic collisions all may modify the local spin state of the components, while still preserving the ensemble-averaged spin.  With these complicated interactions between the two components and within each component, it is often difficult to obtain readily interpretable results in the commingled normal-superfluid regime.  An exception to this is the phenomenon of enhanced domain formation, in which the cooperative nature of normal-superfluid spin interaction is directly manifest.  In this Letter we describe the mechanism for enhanced domain formation and present data demonstrating the effect.

Since the achievement of Bose-Einstein condensation in dilute atomic gases, the interactions between condensates and normal components have been studied in great detail theoretically \cite{finiteTtheory} and to a lesser extent experimentally \cite{finiteTexp}.  An extra complication arises when the gas is comprised of multiple spin levels. Previous work with spinor condensates has primarily been concerned with the energetic and hydrodynamic properties of interpenetrating states with different longitudinal spin population, \emph{i.e.}~studies of spin-domain formation in mostly-pure condensates \cite{ho1998,hall1998,stenger1998}.  There has been a small amount of work \cite{hall1999} devoted to studying properties of nearly pure spinor condensates involving their transverse spin, or internal coherence.  In the fully nondegenerate limit, collective spin wave behavior of a nondegenerate ensemble of ultra-cold atoms has been studied \cite{laloe1982,oldswtheory,bigelow1989,lewando2002,mcguirk2002,swtheory}.  In this work, we study the regime in between nondegenerate spin waves and pure condensate spin domain formation \cite{nikuni2003}.  We present evidence of enhanced domain formation due to spin locking in a partially-condensed system.  Spin locking is the coherent spin dynamics which occur simultaneously and equivalently in both the normal component and the condensate.  Exchange collisions act to constrain the spin of the condensate to that of the normal component, thus locally locking together the spins of the two components.  The result of this process is that spin domains can form in a condensate immersed in a normal gas up to six times faster than they do in a nearly pure condensate \cite{hydrogen}.

\begin{figure}
\leavevmode
\epsfxsize=3.375in
\epsffile{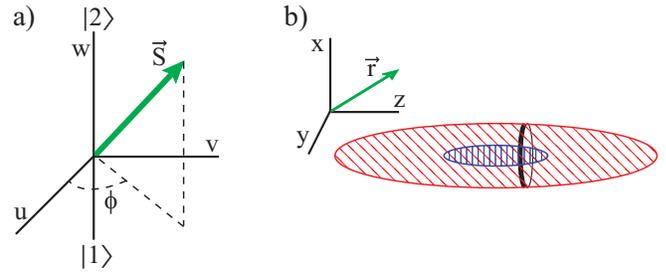} 
\caption{\label{fig:bloch} (Color online) a) Pseudospin coordinates, showing the vector $\vec{S}(\vec{r})$ and transverse phase angle $\phi$.  $\vec{S}(\vec{r}) = \pm \frac{1}{2} \hat{w}$ indicates that the spin population locally is in the pure $|2\rangle$ state and $|1\rangle$ state respectively, and $\phi$ is the phase of the internal coherence.  b) Real space coordinates.  The actual aspect ratio of the normal cloud (angle hatch) and the embedded condensate cloud (vertical hatch) are a factor of six more elongated than shown in this figure. The high collision rate and rapid oscillation frequency in the radial $(x-y)$ directions ensure that the normal and condensate spins are uniform and aligned with each other within each circular cross-section.  We follow the evolution of $\vec{S}$ in time and displacement along the axial direction $z$.}
\end{figure}

The spinor condensate used in this work consists of two sublevels within the $^{87}$Rb groundstate hyperfine manifold with identical magnetic moments, which form a pseudospin doublet that can be magnetically trapped \cite{states}.  We use the framework of the Bloch sphere to describe the spinor \cite{allen1975}.  Fig.~\ref{fig:bloch} indicates the coordinates used in this work.

\begin{figure*}
\leavevmode
\epsfxsize=7in
\epsffile{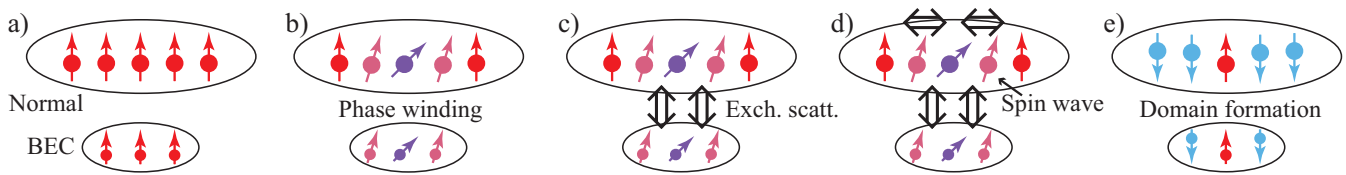} 
\caption{\label{fig:cartoon} (Color online) Schematic explanation of how normal-superfluid spin locking leads to rapid domain formation.  In the sketches, the normal and condensate components are spatially offset for clarity
a) Immediately after a $\pi/2$ pulse, the spin is aligned all along the sample in both components.  b) Mean field effects within each component, and between the components, cause density-dependent evolution of $\phi$, the spin angle in the $u-v$ plane. Although the density is much higher in the condensate, c) exchange scattering keeps the condensate and normal component spins locked at each location along the axis.  d) The axial gradients in $\phi$ launch large-amplitude spin waves through the normal component.  Exchange collisions continue to ensure that the condensate and normal component spin fields are locally aligned so that e) spin domains forms rapidly and simultaneously in both components.}
\end{figure*}

At the low temperatures of this experiment, all inter-particle interactions are of the spherically symmetric, s-wave type.  One can think of these collisions as leading to three primary effects:

1. Momentum-changing elastic collisions: these may occur between two normal atoms, or between a normal and a condensate atom.  Over relatively long time scales these collisions enforce thermal equilibrium throughout the sample.  They are not particularly relevant to the phenomena described in this paper, except in that they are necessary to generate the thermally-equilibrated state that exists before the measurements begin.

2. Density-dependent mean field potentials: the real parts of the forward and backward scattering amplitudes of the elastic collisions lead to a mean field energy proportional to density. Because the scattering lengths $a_{11}$ and $a_{22}$ (where $a_{ij}$ is the scattering length between spin states $|i\rangle$ and $|j\rangle$) differ by about 5\%, the local density of atoms contributes to the energy splitting between the two spin projections.  Therefore $\dot{\phi}$ \cite{phidot} is not uniform across the sample but depends on the local density \cite{interspecies}.

3. Spin-rotation effect: When two colliding atoms are not perfectly aligned, there is a spin-rotation effect that arises from the coherent interference between a back-scattered event and an unscattered event.  As two atoms pass by each other, each atom's spin experiences a small rotation about the vector of their total spin.  This effect occurs between two normal atoms \cite{laloe1982}, and between a normal and condensate atoms \cite{nikuni2003}, but not between two condensate atoms, since condensate atoms are all in the same motional state.  These collisions preserve total spin  but can alter the spatial spin distribution.

The most readily accessible tool for studies of spinor gases, whether condensed or otherwise, is spin domain formation.  In a pure condensate, for instance, the interaction energies of the two states are slightly different, and therefore it is energetically favorable in trap for the condensate to form spin domains (if the intraspecies scattering lengths are not be significantly greater than the interspecies scattering length) \cite{interspecies}.  This was shown to be the case for $^{87}$Rb \cite{hall1998}.  In a pure condensate the driving mechanism for the spin transport comes from the spatial inhomogeneity of the relative mean field shift of the two spin states.  This produces a spatial gradient in the phase, $\phi$.  In a condensate, which is a single wavefunction, a gradient in the relative phase of the spin states means perforce there is a relative velocity between the two states, given by $u(z) = (\hbar/m) \nabla \phi(z)$ \cite{dalfovo1999}.  However, as described below, this mechanism of spin transport is relatively unimportant in condensates when a significant normal component is present.

The spin dynamics of condensates with large normal components are instead dominated by spin waves in the normal component.  Condensates themselves do not support spin waves due to the absence of the spin rotation effect described above, but spin locking to the spin wave in the normal component vastly accelerates domain formation by using exchange scattering as a mechanism for spin transport.  This process is illustrated in Fig.~\ref{fig:cartoon}.  With all the spins in the ensemble initially rotated uniformly into the \emph{u-v} plane, $\phi$ begins to evolve nonuniformly because of the spatial inhomogeneity in the potential, leading to a spatial phase gradient, which in turn produces spin waves in the normal component \cite{mcguirk2002,swtheory}.  Spin waves occur as a number of spin rotating collisions between normal component atoms with slightly different spins (due to the inhomogeneity of the potential) produce a macroscopic spin current, which drives spatio-temporal spin oscillations.  For the large density inhomogeneities typical in this work, spin waves are strongly driven outside of the linear regime, leading to oscillation frequencies dependent on the driving amplitude \cite{laloe2003}.  During the progression of the spin wave, the condensate and normal components undergo exchange scattering that locally equilibrates their spins, thereby entraining the condensate spin dynamics in the normal component spin wave.  The highly overdriven spin wave causes rotation of the normal component spin vectors out of the \emph{u-v} plane on an even faster time scale than spin waves above $T_{\mbox{\scriptsize{c}}}$.  Spin locking of the condensate to the normal component via exchange scattering causes the condensate to form spin domains on the exact same timescale.

\begin{figure}
\leavevmode
\epsfxsize=3.375in
\epsffile{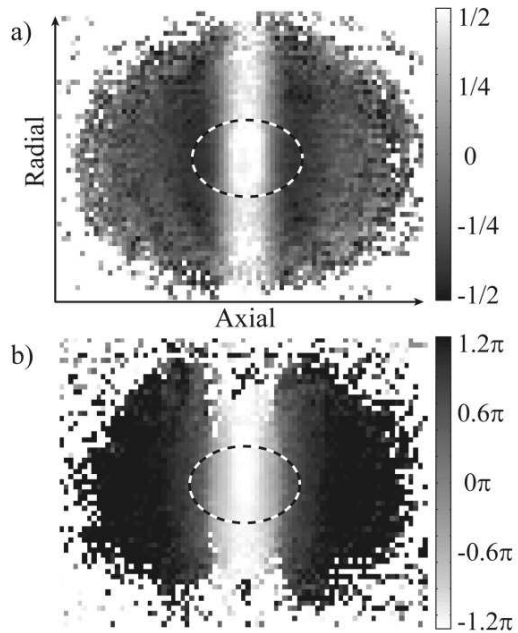}
\caption{\label{fig:blowup} a) Longitudinal spin profile at $T/ T_{\mbox{\scriptsize{c}}} = 0.8$ after a 50~ms evolution time: $\langle S_{\mbox{\scriptsize{w}}}(\vec{r}) \rangle = (N_2(\vec{r})-N_1(\vec{r}))/ (N_2(\vec{r})+N_1(\vec{r}))$, calculated from the measured $|1\rangle$ and $|2\rangle$ state populations.  White and black shading represent $\langle S_{\mbox{\scriptsize{w}}}(\vec{r}) \rangle \approx \pm \frac{1}{2}$ respectively (atoms predominantly in $|2\rangle$ and $|1\rangle$ respectively).  The white areas near the edges of each image are a mask for areas where there is not enough signal to determine accurately $\langle S_{\mbox{\scriptsize{w}}} (\vec{r}) \rangle$.  b) Corresponding spin profile showing the transverse phase angle $\phi$, obtained from Fourier transforms of Ramsey fringes at each two-dimensional spatial bin.  The dotted lines show the Thomas-Fermi radius of the condensate.  The spins of the condensate and normal component are well aligned as seen by the radial homogeneity; there is no signature of the condensate in the spin profile, despite the tendency for $\dot{\phi}$ to be different between condensate and normal component due to differences in mean-field effects.}
\end{figure}

The experiment begins by first precooling $^{87}$Rb atoms in a magneto-optical trap, transferring them to a hybrid Ioffe-Pritchard magnetic trap (frequencies are \{7, 230, 230\}~Hz) via a servo-controlled linear track, and cooling them to degeneracy by forced radio-frequency evaporation \cite{lewando2003a}.  For this work, the final temperature is adjusted to various values ranging from above $T_{\mbox{\scriptsize{c}}}$ to $T/T_{\mbox{\scriptsize{c}}} < 0.3$ \cite{condfraction}.  For all experiments the condensate number was kept constant at $\sim 6.5 \times 10^4$, a relatively small number chosen to limit dipolar relaxation in the upper spin state.  Each experimental cycle is begun with an ensemble in $|1\rangle$, $S(\vec{r}) = -\frac{1}{2}\hat{w}$.  A $\pi/2$ pulse then rotates all spin vectors to lie entirely in the transverse \emph{u-v} plane, creating an equal coherent superposition of $|1\rangle$ and $|2\rangle$ \cite{spinmanip}.  The spinor system is allowed to evolve, and the resulting spatial spin distribution is probed.  The longitudinal spin component $\langle S_{\mbox{\scriptsize{w}}}(\vec{r}) \rangle$ is measured by independently imaging the $|1\rangle$ and $|2\rangle$ components of the superposition.  The transverse spin components are measured by applying a second $\pi/2$ pulse to create Ramsey fringes and measure $\phi$ \cite{mcguirk2002}.  The imaging is accomplished after expansion by destructive absorptive imaging.  Fig.~\ref{fig:blowup} shows two-dimensional spin reconstructions for a partially-condensed ensemble.  Because the condensate component is significantly more compact than the normal component, it is possible to distinguish the two components by their location in the radial direction.  One can thus separately observe the spin dynamics in the condensate and in the normal gas.

\begin{figure*}
\leavevmode
\epsfxsize=7in
\epsffile{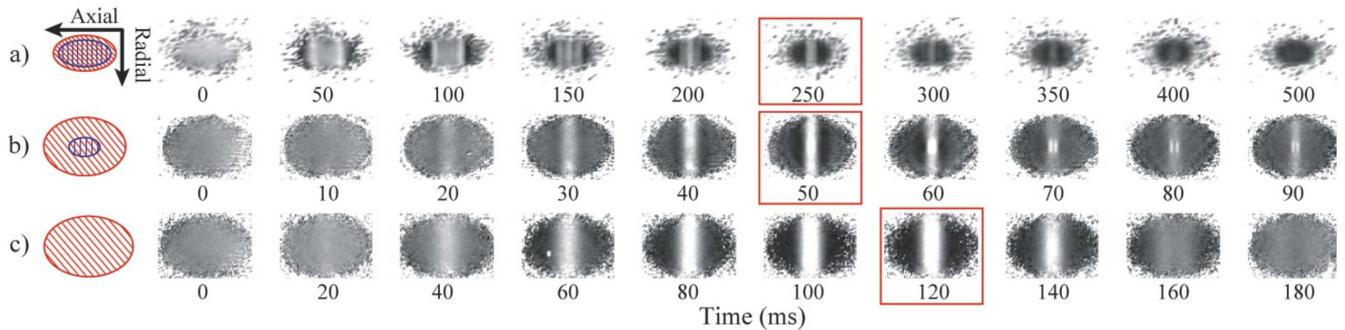}
\caption{\label{fig:strip} (Color online) Temporal evolution of $\langle S_{\mbox{\scriptsize{w}}}(\vec{r}) \rangle$ for $T/ T_{\mbox{\scriptsize{c}}} =$ a) 0.3, b) 0.8, and c) 1.1.  The shading scale for each image is identical to Fig.~\ref{fig:blowup}a.  The first image of each row shows the fitted size of the condensate (vertical hatch) and the normal component (angle hatch), where the sizes are defined by the Thomas-Fermi radius and the $\sim e^{-2}$ radius respectively.  In b), the condensate occupies the center third of each image, while the condensate dominates the image in a) and is not present in c).  The time scale to reach the maximum spin domain formation (boxed image in each row) is highly accelerated when there is a significant normal component present as in b) and is even faster than in c) due to the extra energy inhomogeneity from the mean field shift of the condensate.  Images are 120~$\mu$m across in a) and 300~$\mu$m in b) and c).}
\end{figure*}

The inter-component dynamics are elucidated in Fig.~\ref{fig:strip}.  Fig.~\ref{fig:strip}a shows domain formation in a mostly pure condensate $(T/ T_{\mbox{\scriptsize{c}}}= 0.3)$.  The $|2\rangle$ state forms a domain in the center of the trap, and the $|1\rangle$ state moves to the outside to minimize energy, as described above.  The time scale for this motion is $\sim 280$~ms for a condensate with number density $\sim 10^{14}$~cm$^{-3}$.  The other end of the temperature range is shown in Fig.~\ref{fig:strip}c where spin waves in a cloud just above degeneracy $(T/T_{\mbox{\scriptsize{c}}} = 1.1)$ are depicted.  The maximum spin rotation out of the \emph{u-v} plane, equivalently the peak of the spin domain formation, occurs at $\sim 120$~ms.

The spin dynamics for a partially-condensed system are strikingly different.  An ensemble at $T/ T_{\mbox{\scriptsize{c}}} = 0.8$ is shown in Fig.~\ref{fig:strip}b.  The condensate inhabits the center third of each image, and the normal component is spread throughout the entire image.  The first point to note is the uniformity of the images in the radial direction, which shows spin locking, \emph{i.e.} $\langle S_{\mbox{\scriptsize{w}}}(\vec{r}) \rangle$ is only a function of the axial position $z$.  Second, in the presence of the extra driving potential caused by the condensate, the normal component spin wave has a frequency approximately twice as high as the less strongly driven nondegenerate case.  The last important feature of this data is that spin domains form in the condensate at the same time as the spin wave rotates the normal-fluid spin out of the $u-v$ plane.  Domain formation occurs almost six times faster when driven by condensate-normal component exchange scattering than without the normal component present.

The formation of spin domains in pure condensates, driven by gradients in mean-field potentials, is an effect that was observed six years ago \cite{hall1998,stenger1998}, and this basic effect is seen again in Fig.~\ref{fig:strip}a.  It is tempting to think of the rapid condensate spin-domain formation seen in Fig.~\ref{fig:strip}b as simply an accelerated version of the same physical process.  For instance, the thermal cloud might be contributing to the axial gradient of the mean field potential, such that $\phi(z)$ in the condensate develops a large gradient more quickly, with a corresponding larger relative superfluid velocity between the two spin projections.  We have ruled out this explanation through a quantitative study of the evolution of $\phi(z)$ in the condensate \cite{mcguirk2003}. The observed $\partial\phi/\partial z$ gives the rate of \emph{potential-driven} relative flow of the two spin projections, while the divergence of this flow, $\partial^2\phi/\partial z^2$, would be proportional to the rate at which the local value of $S_w$ evolves towards a pure $|1\rangle$ projection or a pure $|2\rangle$ projection.  Indeed, this analysis can account within a factor of $1.5$ for the rate of spin-domain formation in the case of the near-pure condensate shown in Fig.~\ref{fig:strip}a. However the potential-drive superfluid flow measured within the condensate for the conditions corresponding to the mixed-case example of Fig.~\ref{fig:strip}b are too small by a factor of 40 to account for the observed spin-domain formation rate $\dot{S}_w$.  Obviously then, the spin dynamics of the condensate, which cannot support its own spin wave, arise predominantly from interactions with the normal cloud spin wave.

In conclusion, we have studied normal-superfluid spinor interactions in a regime in which the (normally highly complex) effects admit a simple interpretation.  To complement the dynamics presented here, a study of the causes and effects of spin decoherence in a partially condensed system will be the topic of a future publication \cite{lewando2003b}.

We acknowledge useful conversations with other members of the JILA BEC collaboration.  This work was supported by the NSF and NIST.


\begin{thebibliography}{Hi!}
\bibitem{states}
The two spin states in question are the hyperfine states $|1\rangle \equiv |F=1, m_{\mbox{\scriptsize{F}}} = -1 \rangle$ and $|2\rangle \equiv |F=2, m_{\mbox{\scriptsize{F}}} = 1 \rangle$ of the $5S_{1/2}$ manifold.  This pseudo-spin doublet can be thought of as a spin-$\frac{1}{2}$ system.
\bibitem{phidot}
The mean field-driven phase evolution of the system is determined from the local density $n$ of the components as follows:
\begin{eqnarray}
\dot{\phi}_{norm} =& \frac{4\pi\hbar (a_{11}-a_{22})}{m}(2 n_{norm} + 2 &n_{cond}) \nonumber \\
\dot{\phi}_{cond} =& \frac{4\pi\hbar (a_{11}-a_{22})}{m}(2 n_{norm} + &n_{cond}) \nonumber
\end{eqnarray}
\end{thebibliography}
\end{document}